\documentclass[12pt]{iopart}

\usepackage{subfigure, iopams, graphicx, upgreek, epstopdf}

\renewcommand{\eref}[1]{Eq.~(\ref{#1})}
\newcommand{\erefs}[1]{Eqs.~(\ref{#1})}
\renewcommand{\fref}[1]{Fig.~\ref{#1}}

\renewcommand{\mat}[1]{#1}
\newcommand{\abs}[1]{\left|#1\right|}

\newcommand{\ie}{i.e.}
\newcommand{\eg}{e.g.}
\newcommand{\force}{\boldsymbol{F}}
\newcommand{\diffn}{\boldsymbol{D}}
\newcommand{\im}[1]{\,\mathrm{Im}\!\left\{#1\right\}}
\newcommand{\re}[1]{\,\mathrm{Re}\!\left\{#1\right\}}
\newcommand{\lre}[1]{\,\mathrm{Re}\!\left\{#1\right.}
\newcommand{\tfrac}[2]{\case{#1}{#2}}

\def \rmd {\mathrm{d}}

\newcommand{\comm}[2]{\Bigl[#1,#2^\dagger\Bigr]}

\begin{document}

\title{Dynamical scattering models in optomechanics: Going beyond the `coupled cavities' model}
\author{Andr\'e Xuereb$^1$, and Peter Domokos$^2$}
\address{$^1$\ Centre for Theoretical Atomic, Molecular and Optical Physics, School of Mathematics and Physics, Queen's University Belfast, Belfast BT7\,1NN, United Kingdom}
\address{$^2$\ Institute for Solid State Physics and Optics, Wigner Research Centre for Physics, H-1525 Budapest P.O.\ Box 49, Hungary}
\eads{\mailto{andre.xuereb@qub.ac.uk, domokos.peter@wigner.mta.hu}}


\begin{abstract}
Recently [A. Xuereb, \emph{et al.},\ Phys.\ Rev.\ Lett.\ \textbf{105}, 013602 (2010)], we calculated the radiation field and the optical forces acting on a moving object inside a general one-dimensional configuration of immobile optical elements. In this article we analyse the forces acting on a semi-transparent mirror in the `membrane-in-the-middle' configuration and compare the results obtained from solving scattering model to those from the coupled cavities model that is often used in cavity optomechanical system. We highlight the departure of this model from the more exact scattering theory when the reflectivity of the moving element drops below about $50$\%.
\end{abstract}

\maketitle

\section{Introduction}
The nontrivial interplay between the external (motional) or internal degrees of freedom of a mobile scatterer coupled to a cavity field, and the cavity field itself has attracted considerable attention over the past two decades. Use has been made of a cavity field to, \eg, interact with single atoms~\cite{Mucke2010,Hetet2011,Specht2011,Ritter2012}, cool atomic motion~\cite{Chan2003,Leibrandt2009,Koch2010,Schleier-Smith2011}, impose spontaneous order through a Dicke phase transition in an ultracold atomic medium~\cite{Black2003,Baumann2010}, couple to the motion of mechanical oscillators~\cite{Schliesser2006,Thompson2008,Kippenberg2008,Groblacher2009a,Aspelmeyer2010}, and even cool this motion down to the vibrational ground state~\cite{Teufel2009,OConnell2010,Chan2011}. The description of these systems, along with most of cavity QED (CQED), follows down the path of the `good cavity' approximation~\cite{Walls1995}:\ the cavity mirrors, be they fixed~\cite{Thompson2008,Power1982} or moving~\cite{Braginsky2002,Groblacher2009a}, bound a region of space such that the electromagnetic field in that region is cut off from the outside world. An alternative approach, based on a scattering picture, is possible. Such an approach can treat very general configurations in one dimension, owing to the power of the transfer matrix method (TMM)~\cite{Deutsch1995,Asboth2008,Xuereb2009b}. In the right limits, the two approaches must of course give rise to the same physics, and indeed they do, even in the case of moving boundaries~\cite{Xuereb2009b}. However, there is no guarantee that one TMM model is always equivalent to the same CQED model; it is the purpose of this paper to use the specific example of a scatterer inside a cavity, \ie, the `membrane-in-the-middle' scheme~\cite{Thompson2008,Jayich2008,Sankey2010} to highlight the differences between these two approaches.\\
Indeed, suppose we have a scatterer, say an atom or a membrane, of reflectivity $r$ ($0\leq\abs{r}\leq1$) placed inside a cavity which, on its own, can be described very well using the `good cavity approximation.' One of two limiting descriptions is generally appropriate for this situation in the CQED picture. (i)~If the scatterer were, \eg, an atom, with $\abs{r}\ll1$, the shape of the mode functions of the field inside the cavity will not change appreciably. In this case it is valid to treat the atom in a weak-coupling approximation and assume that it essentially couples to the unperturbed cavity field.\footnote{ This `weak coupling' criterion is not related to the so-called strong coupling condition of CQED, which refers to the regime in which the internal coherent atom--light coupling leads to a dynamics on a time scale shorter than the characteristic decay time of the dissipation processes. This kind of strong coupling can be achieved without distorting spatially the empty cavity mode functions of the radiation field.} (ii)~On the other hand, if the scatterer were a good mirror, with $\abs{r}$ approaching $1$, this description is no longer valid. Not only does the mirror perturb the shape of the cavity field, but in the good-cavity approximation it \emph{defines} two new modes that communicate by tunnelling of photons through the good mirror. This simple example shows the power of the TMM approach:\ the same TMM model is valid for both situations, and indeed for any situation in between, including absorbing scatterers, with the value of the \emph{polarisability} $\zeta$ of the scatterer determining which of the two situations is being described.
\par
There exists a further, and more fundamental, difference between the two approaches. The TMM deals with moving boundary conditions in a way that goes beyond merely having a dynamically-changing detuning. Indeed, the mode functions themselves in the TMM change dynamically. The implications of such a dynamical situation will not be a concern in the following, and we refer the reader to the recent work by Cheung and Law~\cite{Cheung2011} for a more thorough discussion of this point.
\par
The remainder of this paper shall be organised as follows. In the next section we will briefly summarise the general solution to the TMM with one moving scatterer~\cite{Xuereb2010b,Xuereb2012a}. The following section will apply this general solution to the study of the `membrane-in-the-middle' model and compare it to the commonly used CQED model~\cite{Jayich2008}, following which we will conclude.

\section{General solution to the TMM with a moving scatterer}\label{sec:Model}
\begin{figure}[t]
 \centering
 \includegraphics[scale=0.7]{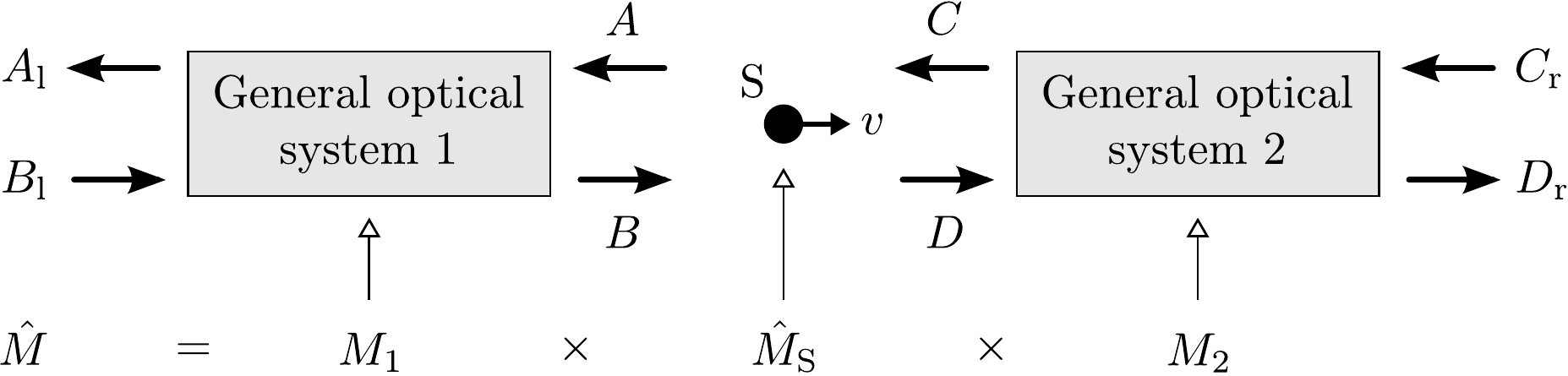}
  \caption{The model we consider in this text, drawn schematically. A scatterer $\mathrm{S}$ interacts with two `general optical systems' in one dimension, composed of immobile linear optical elements, one to either side. $\mathrm{S}$ and these two systems are each represented by a $2\times 2$ matrix.}
 \label{fig:Schematic}
\end{figure}
\subsection{Force acting on the moving scatterer}
Consider the generic situation sketched in \fref{fig:Schematic}. Within the TMM, every scatterer in the situation is represented by a $2\times2$ matrix $M$. Free-space propagation at a wavenumber $k$ is represented by
\begin{equation}
 \mat{M}(k)=\left[\begin{array}{cc}
 e^{ikx}&0\\
 0&e^{-ikx}
 \end{array}\right]\,.
\end{equation} For a static scatterer, $M$ is related simply to the amplitude reflectivity $r$ and transmissivity $t$ of the scatterer, via its polarisability $\zeta$, which may depend on $k$:
\begin{equation}
 \mat{M}(k)=\left[\begin{array}{cc}
 1+i\zeta&i\zeta\\
 -i\zeta&1-i\zeta
 \end{array}\right]=\frac{1}{t}\left[\begin{array}{cc}
 t^2-r^2&r\\
 -r&1
 \end{array}\right]\,.
\end{equation}
Static scatterers do not change the frequency of transmitted and reflected light. A moving scatterer, however, Doppler-shifts reflected light, and we represent this process by transforming $M$ into a frequency-dependent operator $\hat{M}$~\cite{Xuereb2009b,Xuereb2010b}.
At first order in the velocity $v$ of the scatterer, this transformation is remarkably simple and we may write down the general solution for the velocity-dependent force acting on the scatterer in closed form~\cite{Xuereb2010b,Xuereb2012a}. In terms of the notation in \fref{fig:Schematic}, we can define
\begin{equation}
 \hat{\mat{M}}=\mat{M}_1\times\hat{\mat{M}}_\mathrm{S}\times \mat{M}_2\equiv\left[\begin{array}{cc}
 \hat{\gamma}&\hat{\alpha}\\
 \hat{\delta}&\hat{\beta}
 \end{array}\right]\ \mathrm{and}\ \,\bigl(\mat{M}_1\bigr)^{-1}\equiv\bigl[\mu_{ij}\bigr]\,,
\end{equation}
as well as the convenient velocity-independent quantities $\alpha_0$, $\alpha_1^{(0)}$, $\alpha_1^{(1)}$, etc., by
\begin{eqnarray}
 \hat{\alpha}\equiv\alpha_0+\frac{v}{c}\Bigl(\alpha_1^{(0)}+\alpha_1^{(1)}\frac{\partial}{\partial k}\Bigr)\,,\ \hat{\beta}\equiv\beta_0+\frac{v}{c}\Bigl(\beta_1^{(0)}+\beta_1^{(1)}\frac{\partial}{\partial k}\Bigr)\,,\\
 \hat{\gamma}\equiv\gamma_0+\frac{v}{c}\Bigl(\gamma_1^{(0)}+\gamma_1^{(1)}\frac{\partial}{\partial k}\Bigr)\,,\ \mathrm{and}\ \,\hat{\delta}\equiv\delta_0+\frac{v}{c}\Bigl(\delta_1^{(0)}+\delta_1^{(1)}\frac{\partial}{\partial k}\Bigr)\,.
\end{eqnarray}
Assuming that the pumping field is monochromatic about some wavenumber $k_0$, $B_\mathrm{l}=B_0\,\delta(k-k_0)$ and $C_\mathrm{r}=C_0\,\delta(k-k_0)$, we can write the field amplitudes $\mathcal{A}=\int A(k)\,\rmd k$ and $\mathcal{B}=\int B(k)\,\rmd k$ which are given, to first order in $v/c$, by:
\begin{eqnarray}
\label{eq:Afield}
\fl\mathcal{A}=\Biggl(\mu_{11}\frac{\alpha_0}{\beta_0}+\mu_{12}+\frac{v}{c}\Biggl\{\frac{\mu_{11}}{\beta_0^2}\Bigl(\alpha_1^{(0)}\beta_0-\alpha_0\beta_1^{(0)}\Bigr)-\frac{1}{\beta_0}\Biggl[\frac{\partial}{\partial k}\frac{\mu_{11}}{\beta_0}\Bigl(\alpha_1^{(1)}\beta_0-\alpha_0\beta_1^{(1)}\Bigr)\Biggr]\Biggr\}\Biggr)B_0\nonumber\\
\fl\qquad+\Biggl(\mu_{11}\frac{\gamma_0\beta_0-\alpha_0\delta_0}{\beta_0}+\frac{v}{c}\Biggl\{\frac{\mu_{11}}{\beta_0^2}\Bigl[\beta_0^2\gamma_1^{(0)}-\alpha_0\beta_0\delta_1^{(0)}-\Bigl(\alpha_1^{(0)}\beta_0-\alpha_0\beta_1^{(0)}\Bigr)\delta_0\Bigr]\nonumber\\
\fl\qquad\qquad-\Biggl[\frac{\partial}{\partial k}\frac{\mu_{11}}{\beta_0}\Bigl(\beta_0\gamma_1^{(1)}-\alpha_0\delta_1^{(1)}\Bigr)\Biggr]+\frac{\delta_0}{\beta_0}\Biggl[\frac{\partial}{\partial k}\frac{\mu_{11}}{\beta_0}\Bigl(\alpha_1^{(1)}\beta_0-\alpha_0\beta_1^{(1)}\Bigr)\Biggr]\Biggr\}\Biggr)C_0\,,
\end{eqnarray}
and
\begin{eqnarray}
\label{eq:Bfield}
\fl\mathcal{B}=\Biggl(\mu_{21}\frac{\alpha_0}{\beta_0}+\mu_{22}+\frac{v}{c}\Biggl\{\frac{\mu_{21}}{\beta_0^2}\Bigl(\alpha_1^{(0)}\beta_0-\alpha_0\beta_1^{(0)}\Bigr)-\frac{1}{\beta_0}\Biggl[\frac{\partial}{\partial k}\frac{\mu_{21}}{\beta_0}\Bigl(\alpha_1^{(1)}\beta_0-\alpha_0\beta_1^{(1)}\Bigr)\Biggr]\Biggr\}\Biggr)B_0\nonumber\\
\fl\qquad+\Biggl(\mu_{21}\frac{\gamma_0\beta_0-\alpha_0\delta_0}{\beta_0}+\frac{v}{c}\Biggl\{\frac{\mu_{21}}{\beta_0^2}\Bigl[\beta_0^2\gamma_1^{(0)}-\alpha_0\beta_0\delta_1^{(0)}-\Bigl(\alpha_1^{(0)}\beta_0-\alpha_0\beta_1^{(0)}\Bigr)\delta_0\Bigr]\nonumber\\
\fl\qquad\qquad-\Biggl[\frac{\partial}{\partial k}\frac{\mu_{21}}{\beta_0}\Bigl(\beta_0\gamma_1^{(1)}-\alpha_0\delta_1^{(1)}\Bigr)\Biggr]+\frac{\delta_0}{\beta_0}\Biggl[\frac{\partial}{\partial k}\frac{\mu_{21}}{\beta_0}\Bigl(\alpha_1^{(1)}\beta_0-\alpha_0\beta_1^{(1)}\Bigr)\Biggr]\Biggr\}\Biggr)C_0\,,
\end{eqnarray}
where the derivatives are all evaluated at $k=k_0$ and act on the frequency-dependent terms arising from free-space propagation or a $k$-dependent polarisability.\\
To obtain these expressions one first solves for $A_\mathrm{l}(k)$ and $D_\mathrm{r}(k)$ in terms $B_0$ and $C_0$, and then substitutes the results into the matrix equations to obtain explicit expressions for $A(k)$ and $B(k)$. Upon noting that these expressions are valid to first order in $v/c$ and that the pumping field is monochromatic, the integrals can easily be performed to yield \erefs{eq:Afield} and~(\ref{eq:Bfield}). For single-sided pumping (\eg, $C_0=0$), these expressions simplify significantly. We shall find it useful to express these results in the form $\mathcal{A}=\mathcal{A}_0+\frac{v}{c}\mathcal{A}_1$ and $\mathcal{B}=\mathcal{B}_0+\frac{v}{c}\mathcal{B}_1$, with $\mathcal{A}_{0,1}$ and $\mathcal{B}_{0,1}$ being independent of $v$. For conciseness, let us now assume that $\zeta$ does not depend on $k$. Then, using the elements of $\hat{\mat{M}}_\mathrm{S}$, we obtain
\begin{eqnarray}
\fl\mathcal{C}=\int C(k)\,\rmd k=\bigl[(1-i\zeta)\mathcal{A}_0-i\zeta\mathcal{B}_0\bigr]+\tfrac{v}{c}\bigl[(1-i\zeta)\mathcal{A}_1+2i\zeta\mathcal{B}_0-i\zeta\mathcal{B}_1\bigr]\,,
\end{eqnarray}
and
\begin{eqnarray}
\fl\mathcal{D}=\int D(k)\,\rmd k=\bigl[i\zeta\mathcal{A}_0+(1+i\zeta)\mathcal{B}_0\bigr]+\tfrac{v}{c}\bigl[2i\zeta\mathcal{A}_0-i\zeta\mathcal{A}_1-(1+i\zeta)\mathcal{B}_1\bigr]\,.
\end{eqnarray}
We denote the velocity-independent parts of $\mathcal{C}$ and $\mathcal{D}$ by $\mathcal{C}_0$ and $\mathcal{D}_0$, respectively. The force acting on the scatterer can be finally written down as $\force=\force_0+\tfrac{v}{c}\force_1$, where
\begin{eqnarray}
\fl\force_0=-2\hbar k_0\Bigl[&\bigl(\abs{\zeta}^2+\im{\zeta}\bigr)\abs{\mathcal{A}_0}^2+\bigl(\abs{\zeta}^2-\im{\zeta}\bigr)\abs{\mathcal{B}_0}^2\nonumber\\
&+2\re{\bigl(\abs{\zeta}^2+i\re{\zeta}\bigr)\mathcal{A}_0\mathcal{B}_0^\ast}\Bigr]\,,
\end{eqnarray}
and
\begin{eqnarray}
\fl\force_1=-4\hbar k_0\Bigl[&\abs{\zeta}^2\bigl(\abs{\mathcal{A}_0}^2-\abs{\mathcal{B}_0}^2\bigr)+\bigl(\abs{\zeta}^2+\im{\zeta}\bigr)\re{\mathcal{A}_0\mathcal{A}_1^\star}-2\im{\zeta}\re{\mathcal{A}_0\mathcal{B}_0^\star}\nonumber\\
&+\bigl(\abs{\zeta}^2-\im{\zeta}\bigr)\re{\mathcal{B}_0\mathcal{B}_1^\star}+\im{\zeta}\re{\mathcal{A}_0\mathcal{B}_1^\star}\nonumber\\
&+\re{\bigl(\abs{\zeta}^2+i\re{\zeta}\bigr)\mathcal{A}_1\mathcal{B}_0^\star}\Bigr]\,;
\end{eqnarray}
the quantity $\rmd\force/\rmd v=c\force_1$ will henceforth be called the `friction coefficient'.

\subsection{Momentum diffusion experienced by the moving scatterer}
The field amplitudes calculated in the previous section related to classical electromagnetic fields. We may now impose a canonical quantisation on these fields~\cite{Xuereb2009b}, promoting each field variable $A$, say, to an operator $\hat{A}$, such that $\langle\hat{A}\rangle=\sqrt{2\epsilon_0S/\bigl(\hbar k_0\bigr)}\,A$, $S$ being the mode cross-sectional area. The only two \emph{a priori} independent modes in our system are the two input modes $\hat{B}_\mathrm{l}$ and $\hat{C}_\mathrm{r}$, whose operators obey the usual bosonic commutation relations
\begin{equation}
\label{eq:BasicComms}
 \comm{\hat{B}_\mathrm{l}}{\hat{B}_\mathrm{l}}=\comm{\hat{C}_\mathrm{r}}{\hat{C}_\mathrm{r}}=1\,,\ \mathrm{and}\ \comm{\hat{B}_\mathrm{l}}{\hat{C}_\mathrm{r}}=0\,.
\end{equation}
The commutation relations between each of the four fields $\hat{A}$, $\hat{B}$, $\hat{C}$, and $\hat{D}$ can then be built up; because $\force$ is correct up to first order in $v/c$ we only need to evaluate expressions to zeroth order in this section. The fluctuations in these fields will lead to a diffusion in momentum-space, quantified by the diffusion coefficient $\diffn$. Another contribution to $\diffn$ is due to lossy scatterers:\ any absorptive scatterer effectively couples the system to a further, `loss,' mode that is independent of the input fields and is necessary to preserve the canonical commutation relations~\cite{Xuereb2009b}. Such loss modes can be included self-consistently into the TMM~\cite{Xuereb2012a}. Putting all of this together we can write
\begin{eqnarray}
\fl\diffn=&\bigl(\hbar k_0\bigr)^2\Bigl(\left|\mathcal{A}_0\right|^2\comm{\hat{A}}{\hat{A}}+\left|\mathcal{B}_0\right|^2\comm{\hat{B}}{\hat{B}}+\left|\mathcal{C}_0\right|^2\comm{\hat{C}}{\hat{C}}+\left|\mathcal{D}_0\right|^2\comm{\hat{D}}{\hat{D}}\nonumber\\
\fl&\qquad\qquad+2\lre{\mathcal{A}_0^\star\mathcal{B}_0\comm{\hat{A}}{\hat{B}}-\mathcal{A}_0^\star\mathcal{C}_0\comm{\hat{A}}{\hat{C}}}-\mathcal{A}_0^\star\mathcal{D}_0\comm{\hat{A}}{\hat{D}}-\mathcal{B}_0^\star\mathcal{C}_0\comm{\hat{B}}{\hat{C}}\nonumber\\
\fl&\qquad\qquad\phantom{2Re\qquad}\left.-\ \mathcal{B}_0^\star\mathcal{D}_0\comm{\hat{B}}{\hat{D}}+\mathcal{C}_0^\star\mathcal{D}_0\comm{\hat{C}}{\hat{D}}\right\}\Bigr)\,.
\end{eqnarray}
Knowledge of $\diffn$ and $\force$ then allows us to obtain the equilibrium temperature to which the scatterer will tend to:
\begin{equation}
 k_\mathrm{B}T=-\diffn/(c\force_1)\,,
\end{equation}
where $k_\mathrm{B}$ is Boltzmann's constant. These quantities, which can thus be fully determined from our scattering model, are some of the more important quantities of interest in optomechanical setups and atom-CQED, and allow us to describe the dynamical behaviour of such systems.

\begin{figure}[t]
 \centering
 \includegraphics[scale=0.7]{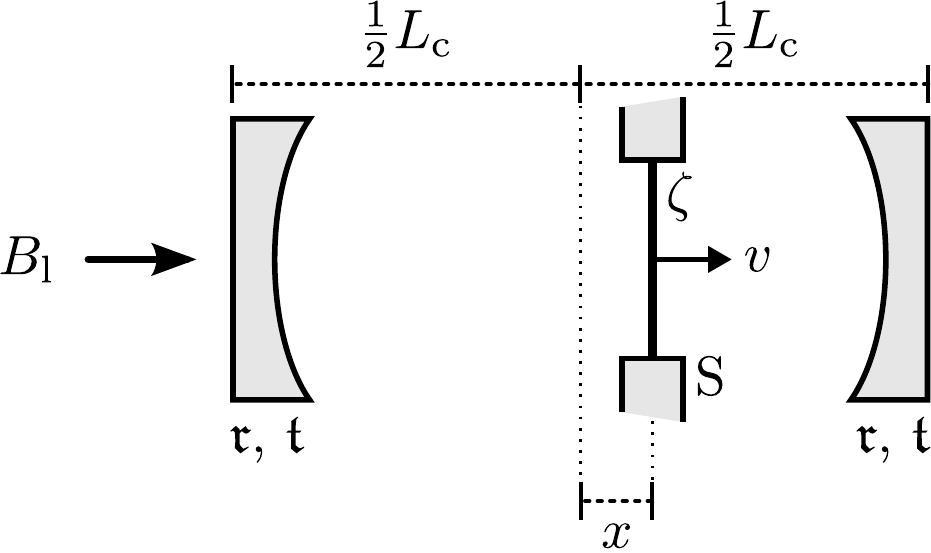}
  \caption{Our model for the `membrane-in-the-middle' geometry:\ the general optical systems in \fref{fig:Schematic} have been replaced by identical mirrors that form a cavity around the moving scatterer. We will only consider situations where $\abs{x}\ll L_\mathrm{c}$.}
 \label{fig:CMC}
\end{figure}
\section{`Membrane-in-the-middle' model}
\begin{figure}[t]
 \centering
 \subfigure[\ $\zeta=-0.100$]{
  \includegraphics[angle=-90,width=0.4\textwidth]{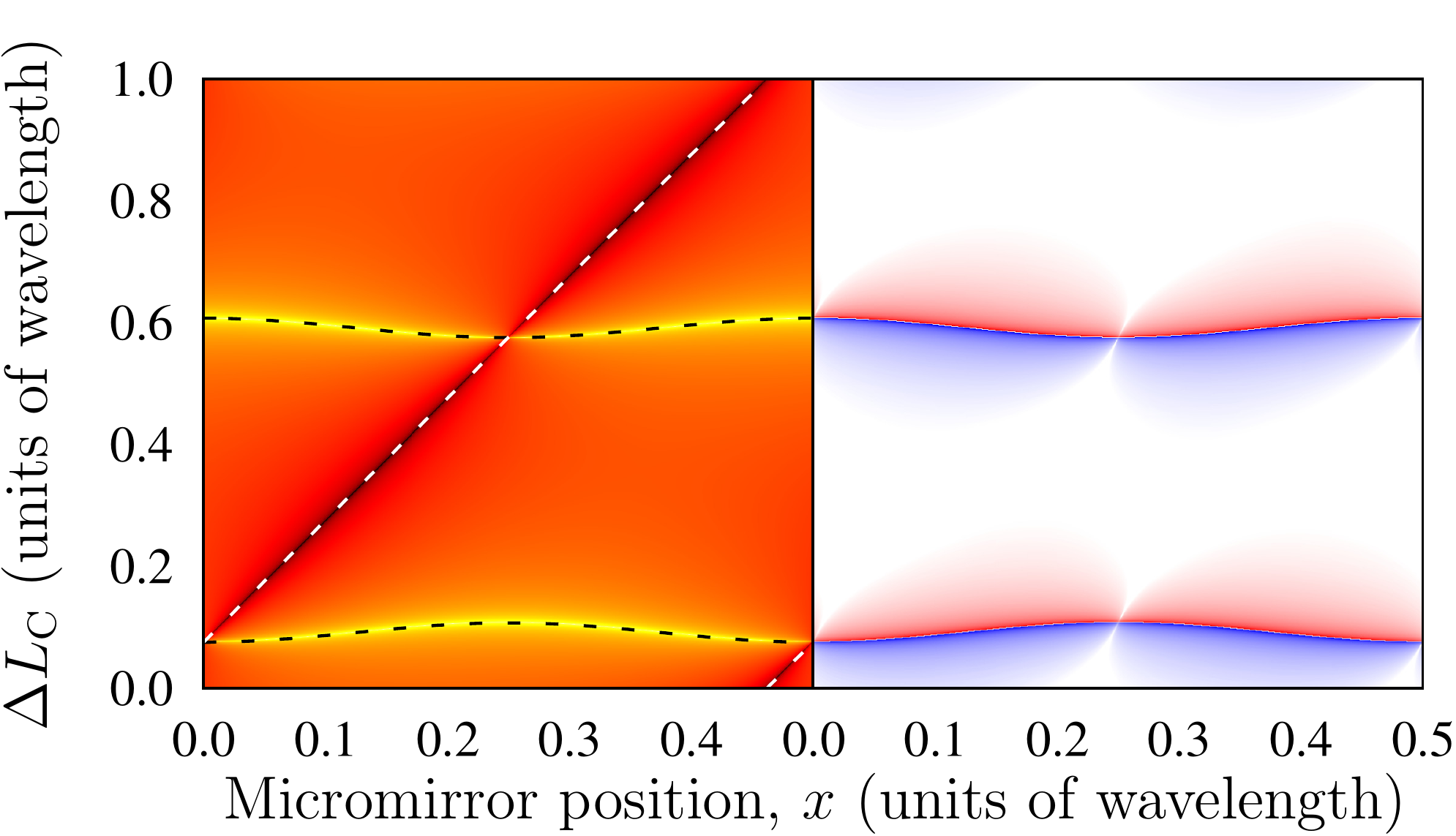}
 }
 \subfigure[\ $\zeta=-0.500$]{
  \includegraphics[angle=-90,width=0.4\textwidth]{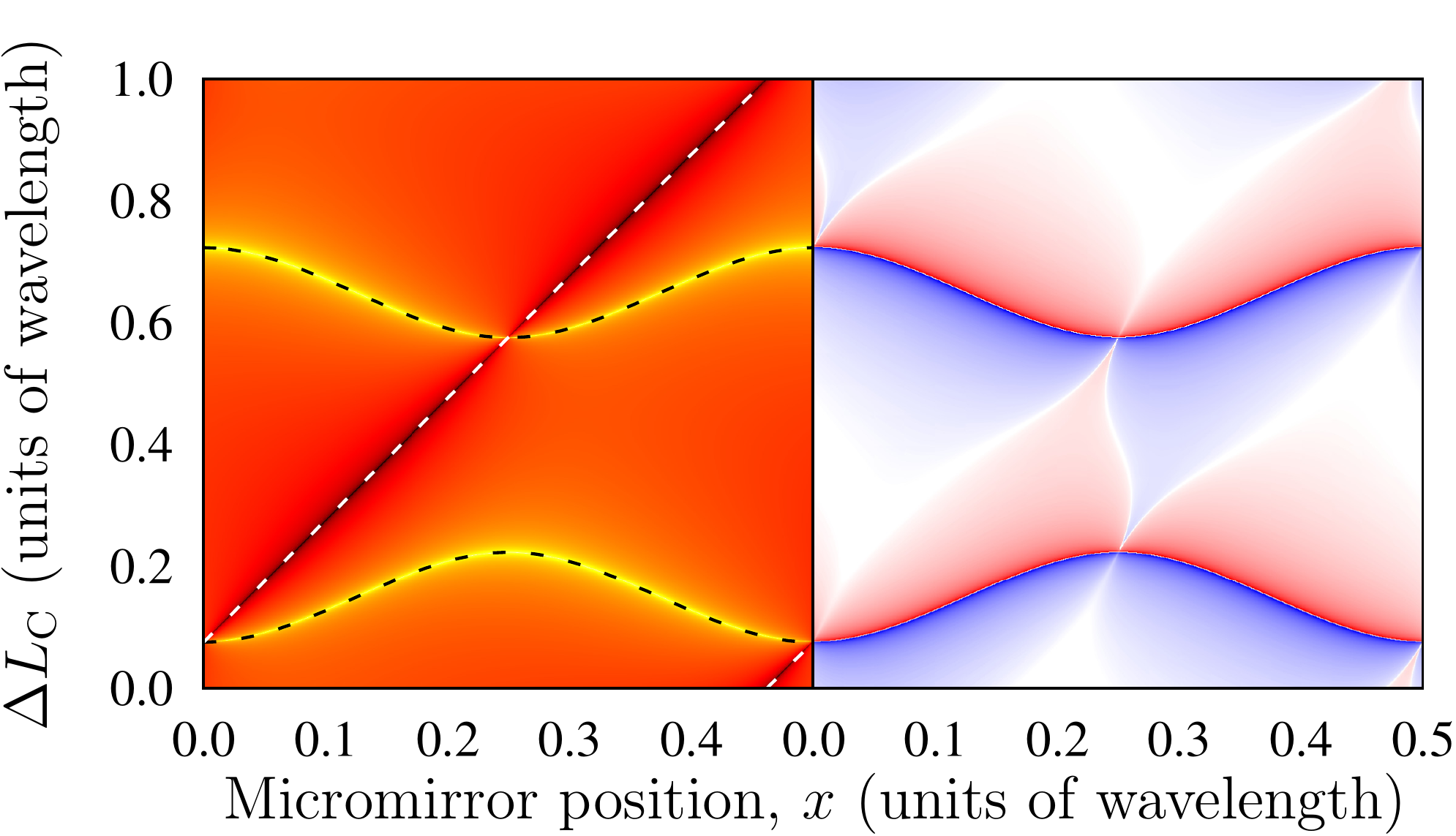}
 }\\
 \subfigure[\ $\zeta=-1.000$]{
  \includegraphics[angle=-90,width=0.4\textwidth]{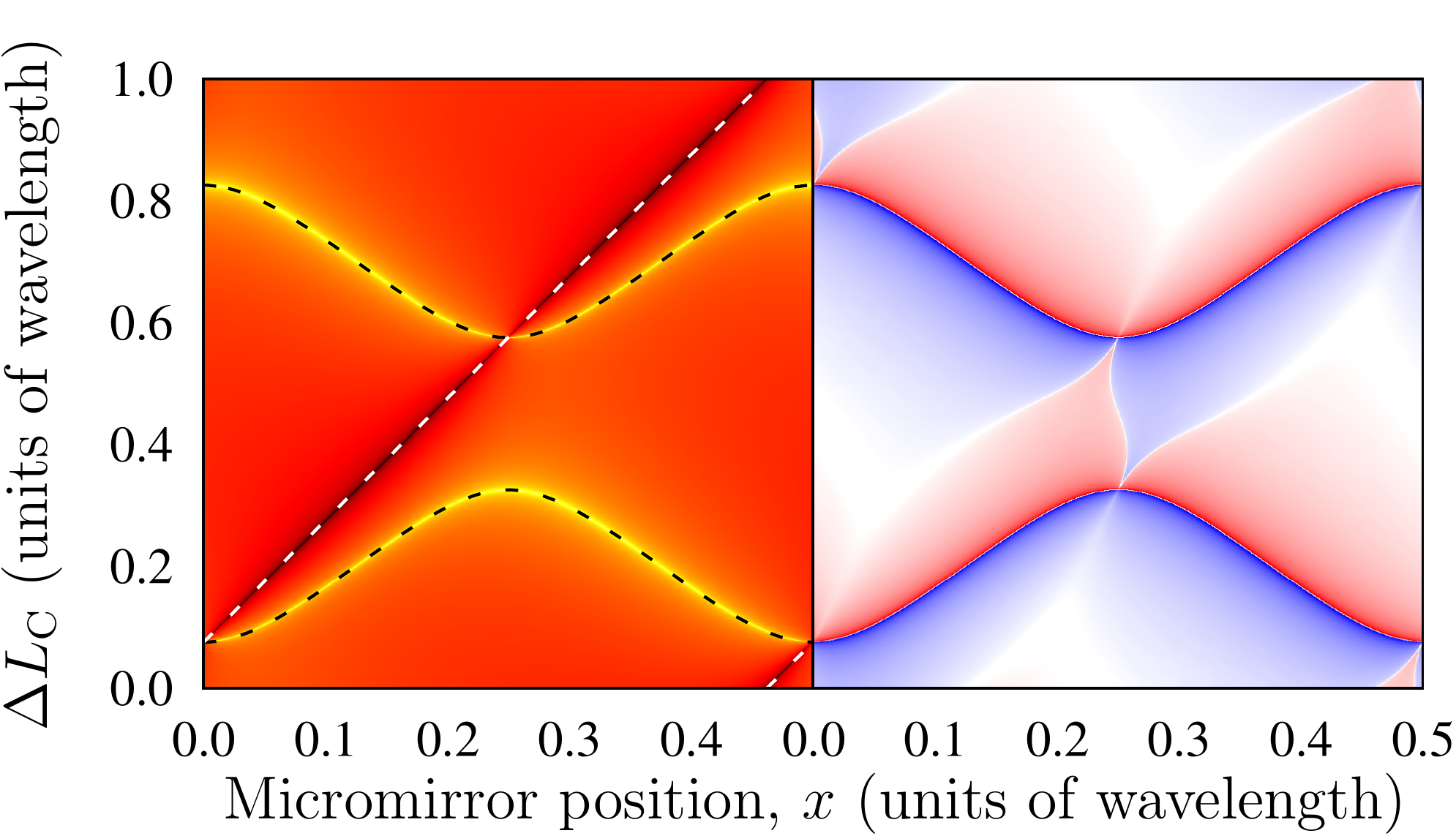}
 }
 \subfigure[\ $\zeta=-2.000$]{
  \includegraphics[angle=-90,width=0.4\textwidth]{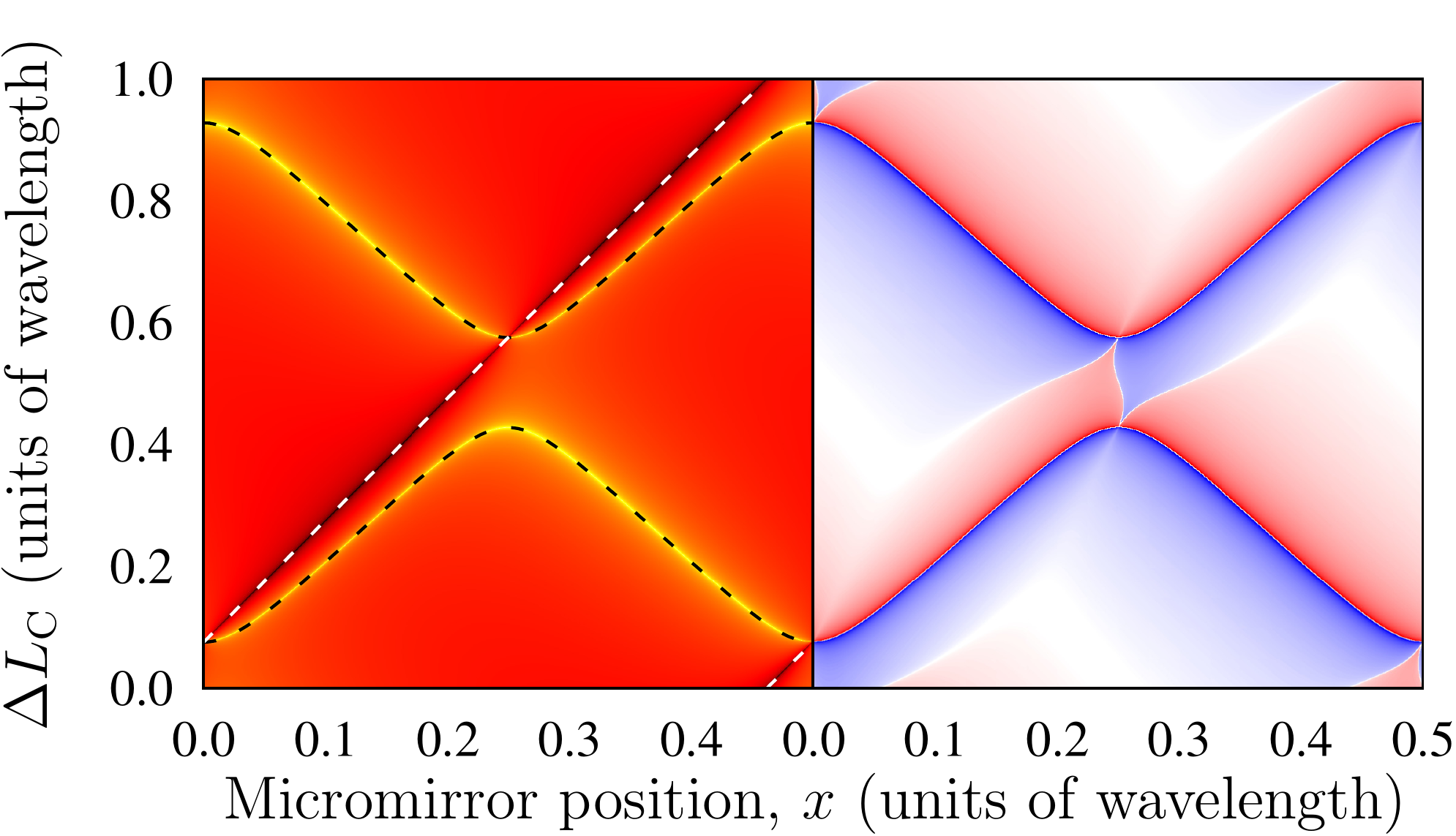}
 }\\
 \subfigure[\ $\zeta=-5.000$]{
  \includegraphics[angle=-90,width=0.4\textwidth]{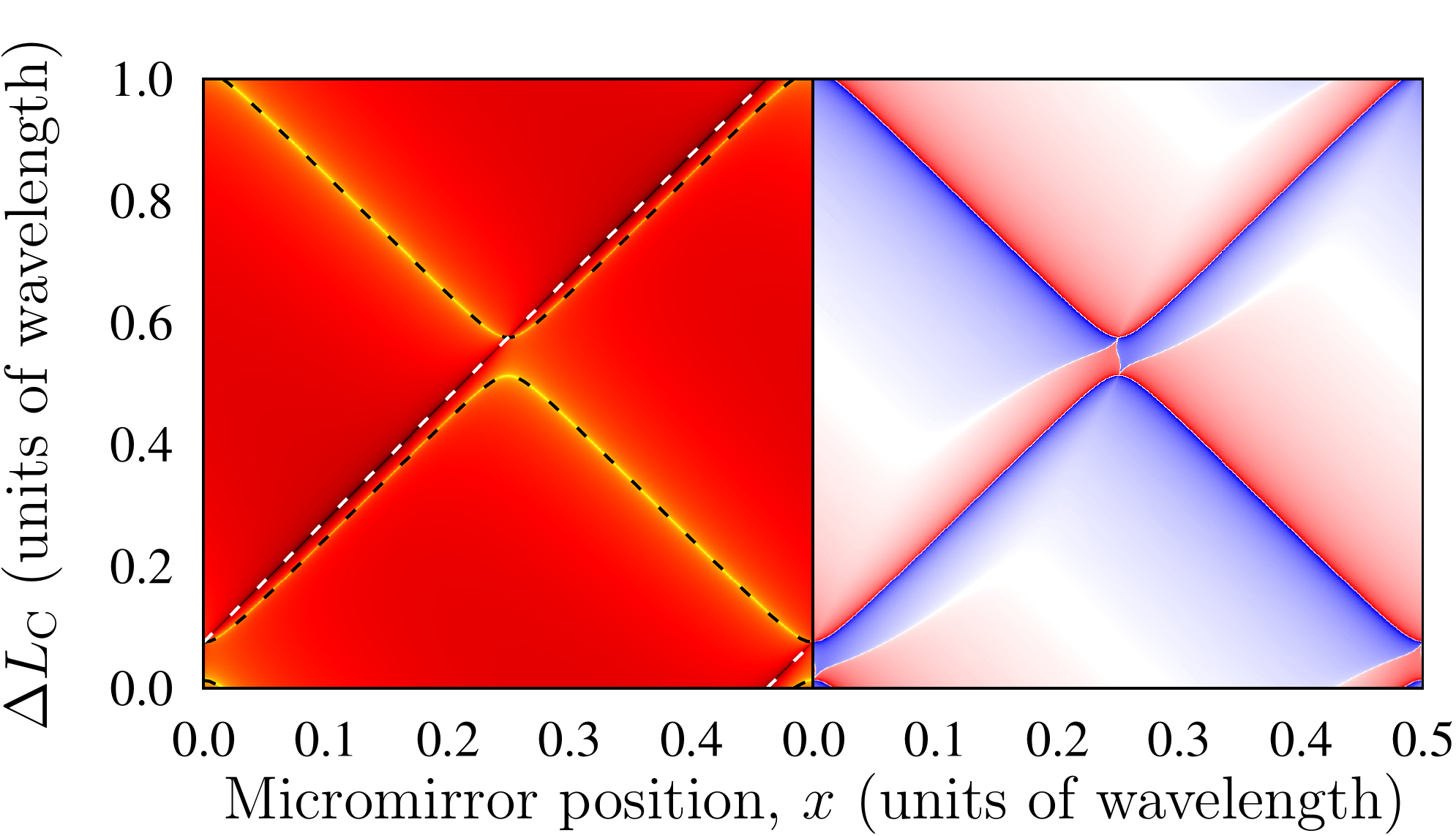}
 }
 \subfigure[\ $\zeta=-10.000$]{
  \includegraphics[angle=-90,width=0.4\textwidth]{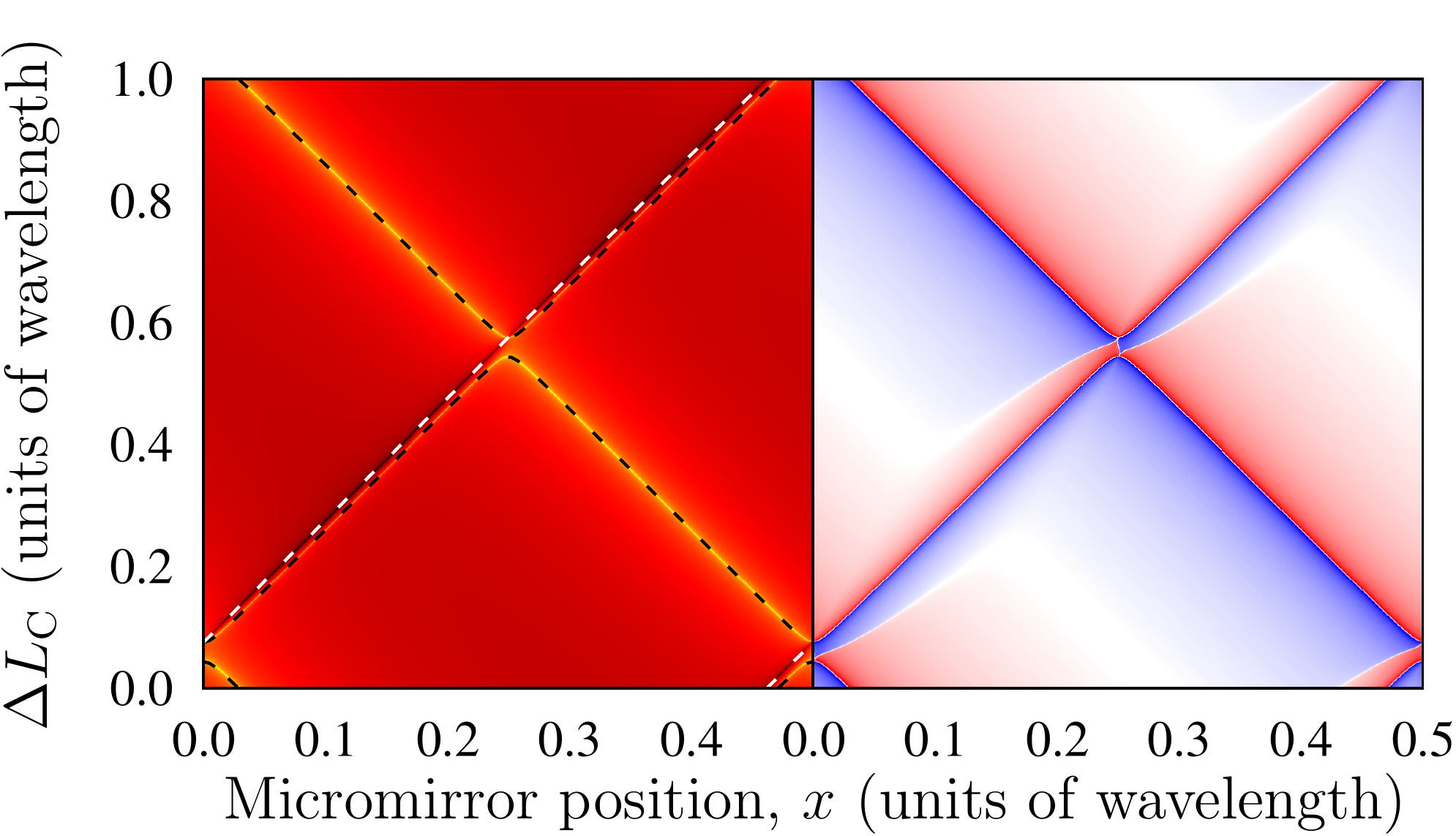}
 }\\
 \subfigure{
  \includegraphics[angle=-90,scale=0.3]{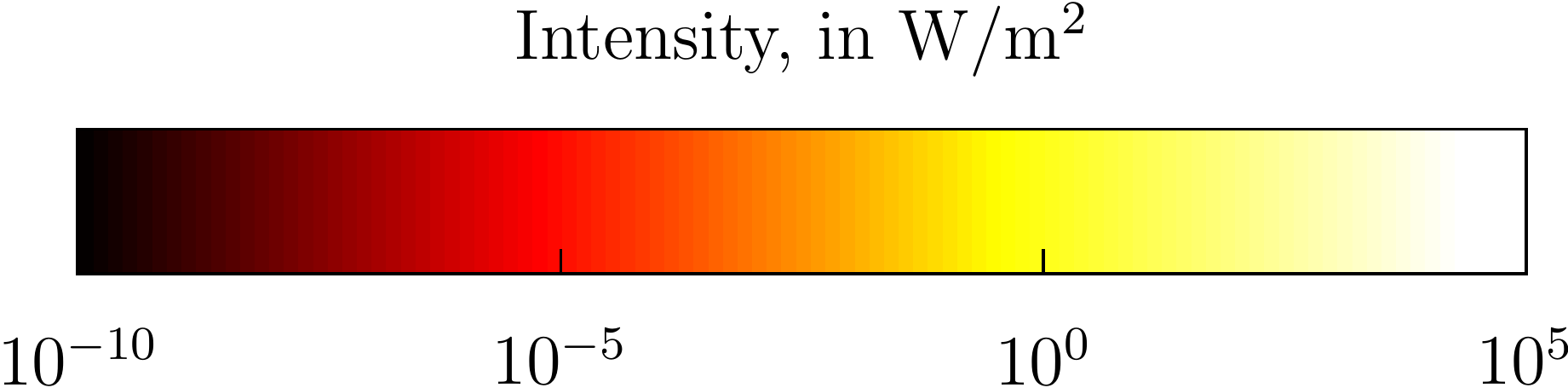}
 }\hspace{1cm}
 \subfigure{
  \includegraphics[angle=-90,scale=0.3]{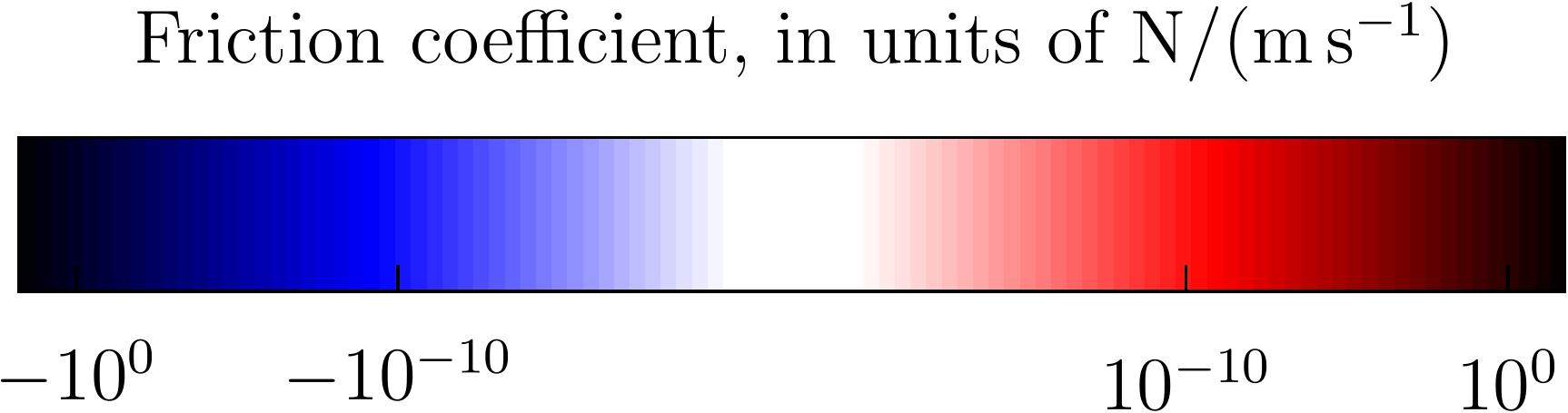}
 }
 \caption{Field intensity (left panels) at and friction coefficient (right panels) acting on the micromirror as the micromirror position ($x$) and cavity length ($L_\mathrm{c}+\Delta L_\mathrm{c}$) are scanned. The subfigures differ only in the polarisability of the mirror, as indicated. The cavity parameters are modelled from Ref.~\cite{Thompson2008}. In the series of left panels, we note the progression from an almost bare cavity situation (a) to a very strong perturbation by the micromirror, leading to avoided crossings (f). The white dashed line traces a cavity node, whereas the black dashed lines [\eref{eq:CavityResonances}] trace the cavity resonances. In the series of right panels, note that the friction coefficient is---as expected---a cooling force (blue) for red cavity detuning and a heating force (red) for blue detuning. The colourbars are on a logarithmic scale and are for $1$\,W of input power.}
 \label{fig:IaF}
\end{figure}

We begin by modelling the system in Ref.~\cite{Thompson2008}:\ a two--mirror Fabry--P\'erot cavity with a micromirror near its centre, operating at a wavelength $\lambda=1064$\,nm and having a length $L_\mathrm{c}=6.7$\,cm, cf.\ \fref{fig:CMC}. The micromirror is modelled by its polarisability $\zeta$ which, in light of the small losses observed in practice, is taken to be real and negative. Whereas the real experimental system corresponds to $\abs{\zeta}\lesssim1$, we allow $\zeta$ to vary freely in our model. The two quantities of interest in this section are the intensity of the field close to the micromirror, and the friction coefficient acting on the micromirror. The former of these gives us knowledge of the resonant frequencies of the cavity and, therefore, of the optomechanical coupling, to all orders, between the cavity field and the micromirror. The latter is useful in optomechanical cooling experiments; the interest here lies in the fact that cooling the motion of a micromirror is one way towards achieving higher sensitivity in metrology applications, most notably in gravitational-wave detectors~\cite{Braginsky2002}, force sensors~\cite{Gavartin2011b}, and magnetometers~\cite{Forstner2012}.
\par
These quantities are summarised in \fref{fig:IaF}, with the left panels showing the intensity at the mirror and the right panels the friction coefficient acting on the mirror. Each subfigure (a)--(f) explores a different value for $\zeta$. For $\abs{\zeta}\ll1$, the cavity field is close to the bare-cavity field; in particular, the cavity resonances are only slightly perturbed by the presence and position of the micromirror. The opposite is true of the $\abs{\zeta}\gg1$ case, where there is coupling between pairs of cavity modes, typified by the avoided crossing in the spectra. The resonance frequencies can be obtained analytically, in the limit of a good bare cavity, as frequency shifts from the bare resonances:
\begin{equation}
\Delta\omega=\frac{c}{L_\mathrm{c}}\tan^{-1}\left\{\frac{\zeta^2\cos(2k_0x)\pm\sqrt{1+\zeta^2\sin^2(2k_0x)}}{\zeta\Bigl[\cos(2k_0x)\mp\sqrt{1+\zeta^2\sin^2(2k_0x)}\Bigr]}\right\}\,,
\label{eq:CavityResonances}
\end{equation}
with $L_\mathrm{c}$ being the length of the cavity, $x$ the position of the micromirror, and $k_0=2\pi/\lambda$ the wavenumber of the light inside the cavity; \eref{eq:CavityResonances} is identical to Eq.~(4) in Ref.~\cite{Jayich2008}. The two sets of solutions to \eref{eq:CavityResonances} are, in the $\zeta\rightarrow 0$ limit, separated by a free spectral range. These cavity resonances, plotted as detuned cavity lengths $\Delta L_\mathrm{c}=\bigl(L_\mathrm{c}/\omega\bigr)\Delta\omega$, are traced by means of the dashed black curves in the left panels of \fref{fig:IaF}. We note that a unit on the vertical axis ($\Delta L_\mathrm{c}=\lambda$) is equal to twice the free-spectral range of the cavity.\\
In the standard optomechanical coupling Hamiltonian, the mirror--field coupling is represented by a term of the form
\begin{equation}
\hat{H}_\mathrm{OM}^{(1)}\sim\hbar\omega^\prime\hat{x}\hat{a}^\dagger\hat{a}\,,
\end{equation}
where $\hat{x}$ the position operator of the mirror, and $\omega^\prime\equiv\partial(\Delta\omega)/\partial x$. $\hat{a}$ is the annihilation operator of the field mode that has the dominant interaction with the micromirror; in the $\abs{\zeta}\rightarrow 0$ limit, these field modes are the bare cavity modes of the whole cavity. However, as $\abs{\zeta}$ increases, the micromirror effectively splits the main cavity into two coupled cavities, giving rise to symmetric and antisymmetric modes, seen as the higher (bright) and lower (dark) branches in \fref{fig:IaF}(f) for $0<x<\lambda/4$; in such cases $a$ is the annihilation operator belonging to one of these eigenmodes. We note that similar behaviour was observed in Ref.~\cite{Jayich2008}.\\
Certain effects, such as mechanical squeezing of the mirror position~\cite{Nunnenkamp2010} and quantum non-demolition measurements on the mirror~\cite{Clerk2010b}, require not \emph{linear coupling} to $\hat{x}$ but \emph{quadratic coupling} to $\hat{x}^2$:
\begin{equation}
\hat{H}_\mathrm{OM}^{(2)}\sim\hbar\omega^{\prime\prime}\hat{x}^2\hat{a}^\dagger\hat{a}\,,
\end{equation}
with $\omega^{\prime\prime}\equiv\partial^2(\Delta\omega)/\partial x^2$. In our notation, we have
\begin{equation}
\label{eq:LinearCoupling}
\omega^\prime=\pm\frac{2k_0c}{L_\mathrm{c}}\frac{\zeta\sin(2k_0x)}{\big[1+\zeta^2\sin^2(2k_0x)\bigr]^{1/2}}\,,
\end{equation}
and
\begin{equation}
\label{eq:QuadraticCoupling}
\omega^{\prime\prime}=\pm\frac{4k_0^2c}{L_\mathrm{c}}\frac{\zeta\cos(2k_0x)}{\big[1+\zeta^2\sin^2(2k_0x)\bigr]^{3/2}}\,.
\end{equation}
\begin{figure}[t]
 \centering
 \subfigure[]{
 \includegraphics[angle=-90,width=0.45\textwidth]{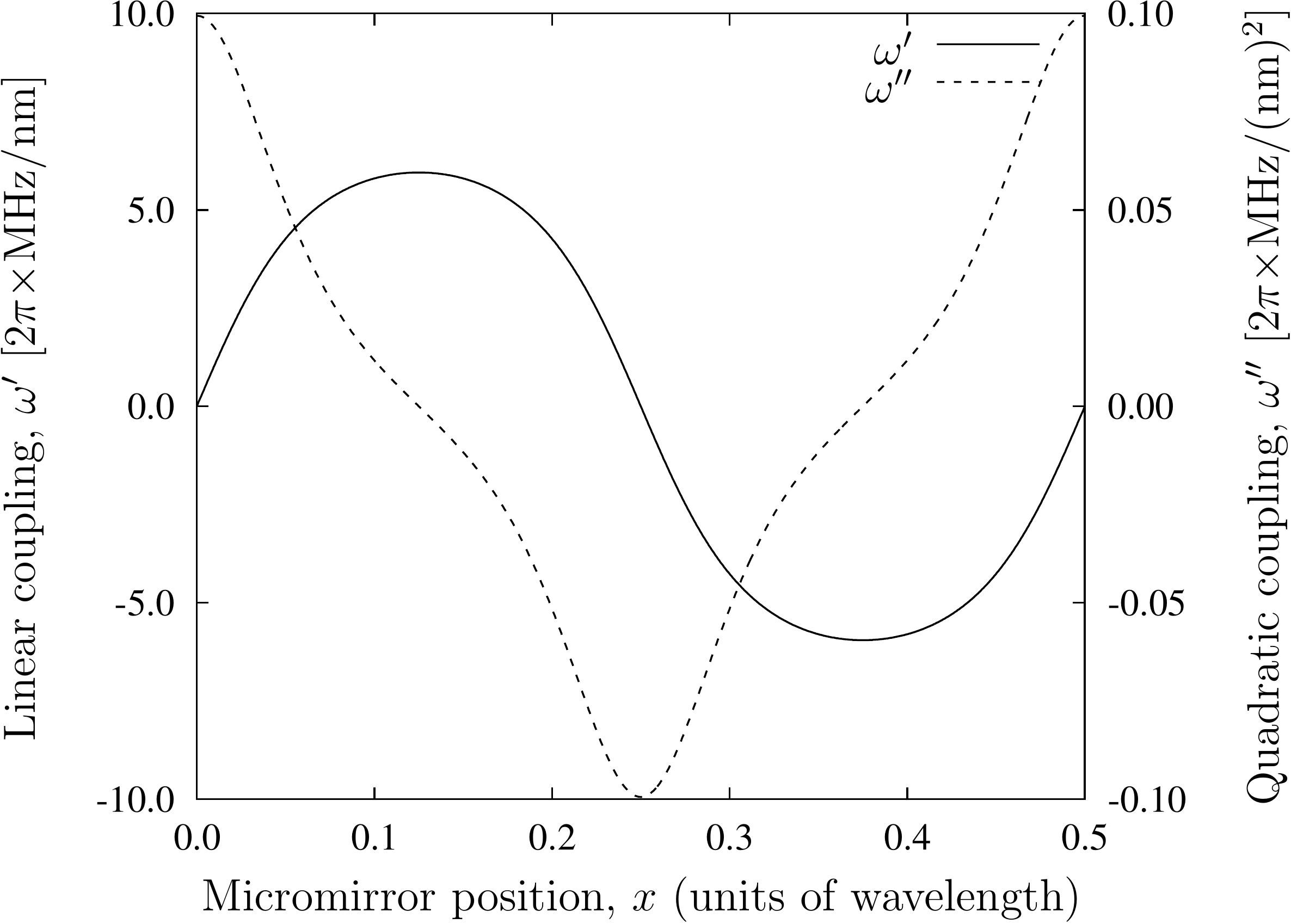}
 }\quad
 \subfigure[]{
 \includegraphics[angle=-90,width=0.45\textwidth]{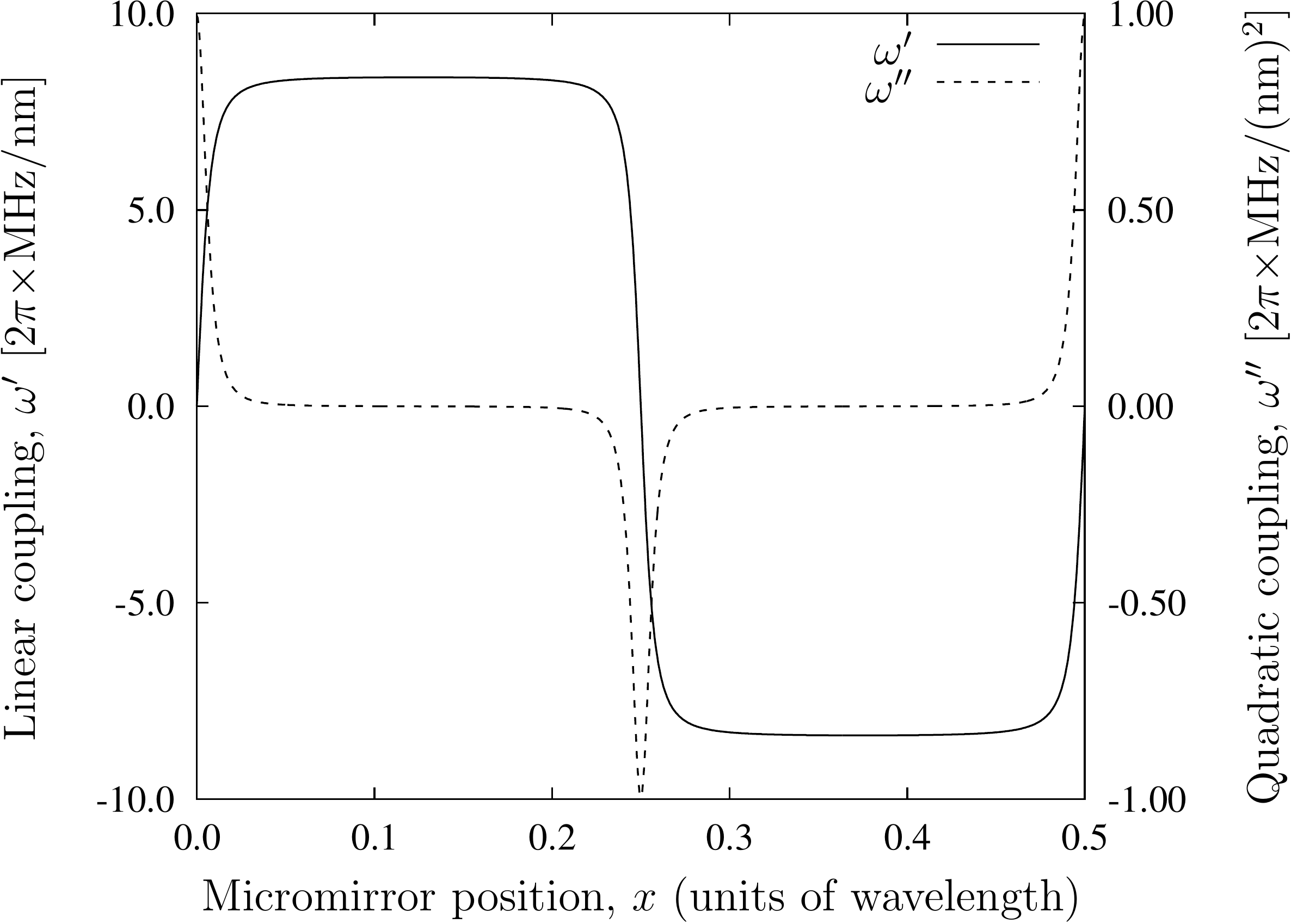}
 }
 \caption{Linear and quadratic optomechanical couplings as a function of mirror position for a very good cavity and for (a)~$\zeta=-1$, and (b)~$\zeta=-10$. In each figure we show the linear (solid curve) and quadratic (dashed curve) couplings, from \erefs{eq:LinearCoupling}~and~(\ref{eq:QuadraticCoupling}). Note that the peak value of $\omega^{\prime\prime}$ is roughly proportional to $\zeta$ whereas $\omega^\prime$ is bounded.}
 \label{fig:Couplings}
\end{figure}
One thing we note immediately is that there is no value for $x$ such that $\omega^\prime=\omega^{\prime\prime}=0$; in other words, the optomechanical coupling is restricted to be linear or quadratic, to lowest order. Higher-order nonlinearities may be achieved by coupling different transverse modes of the cavity (see, e.g., the experimental results in Ref.~\cite{Sankey2010}) but are overwhelmed by the linear or quadratic couplings in a single-transverse-mode cavity. Moreover, the linear coupling $\omega^\prime$ is bounded in the $\zeta\rightarrow\infty$ limit:
\begin{equation}
\abs{\omega^\prime}\leq\frac{2k_0c}{L_\mathrm{c}}\approx 2\pi\times8.42\,\mathrm{MHz/nm}\,,
\end{equation}
with the numeric value corresponding to our parameters. In the same limit, $\omega^{\prime\prime}$ exhibits resonant behaviour (see \fref{fig:Couplings}), indicative of avoided crossings in the spectrum, peaking at a value of:
\begin{equation}
\abs{\omega^{\prime\prime}}\rightarrow\frac{4k_0^2c}{L_\mathrm{c}}\abs{\zeta}\approx 2\pi\times 0.10\,\abs{\zeta}\,\mathrm{MHz/(nm)}^2\,.
\end{equation}
We plot the lower ($\pm\rightarrow-$) branches of \erefs{eq:LinearCoupling} and~(\ref{eq:QuadraticCoupling}) in \fref{fig:Couplings} for two values for $\zeta$:\ $\zeta=-1$, representative of realistic micromirrors, and $\zeta=-10$, representative of a highly reflective micromirror. These correspond to cases (e) and (f) in \fref{fig:IaF}, respectively. Coupling between the pairs of modes is not very strong for the $\zeta=-1$ case; this is manifested by means of the smooth variation with $x$ of $\omega^\prime$ and $\omega^{\prime\prime}$ in \fref{fig:Couplings}(a). The second case shows strong signs of the avoided crossing behaviour seen in \fref{fig:IaF}(f), with $\omega^\prime$ no longer behaving smoothly and $\omega^{\prime\prime}$ acquiring a resonance-like character. Note that, independently of the magnitude of $\zeta$, the strongest quadratic coupling always occurs at the points where $\omega^\prime=0$.

\begin{figure}
 \centering
 \subfigure[\ $\zeta=-0.100$]{
  \includegraphics[angle=-90,width=0.4\textwidth]{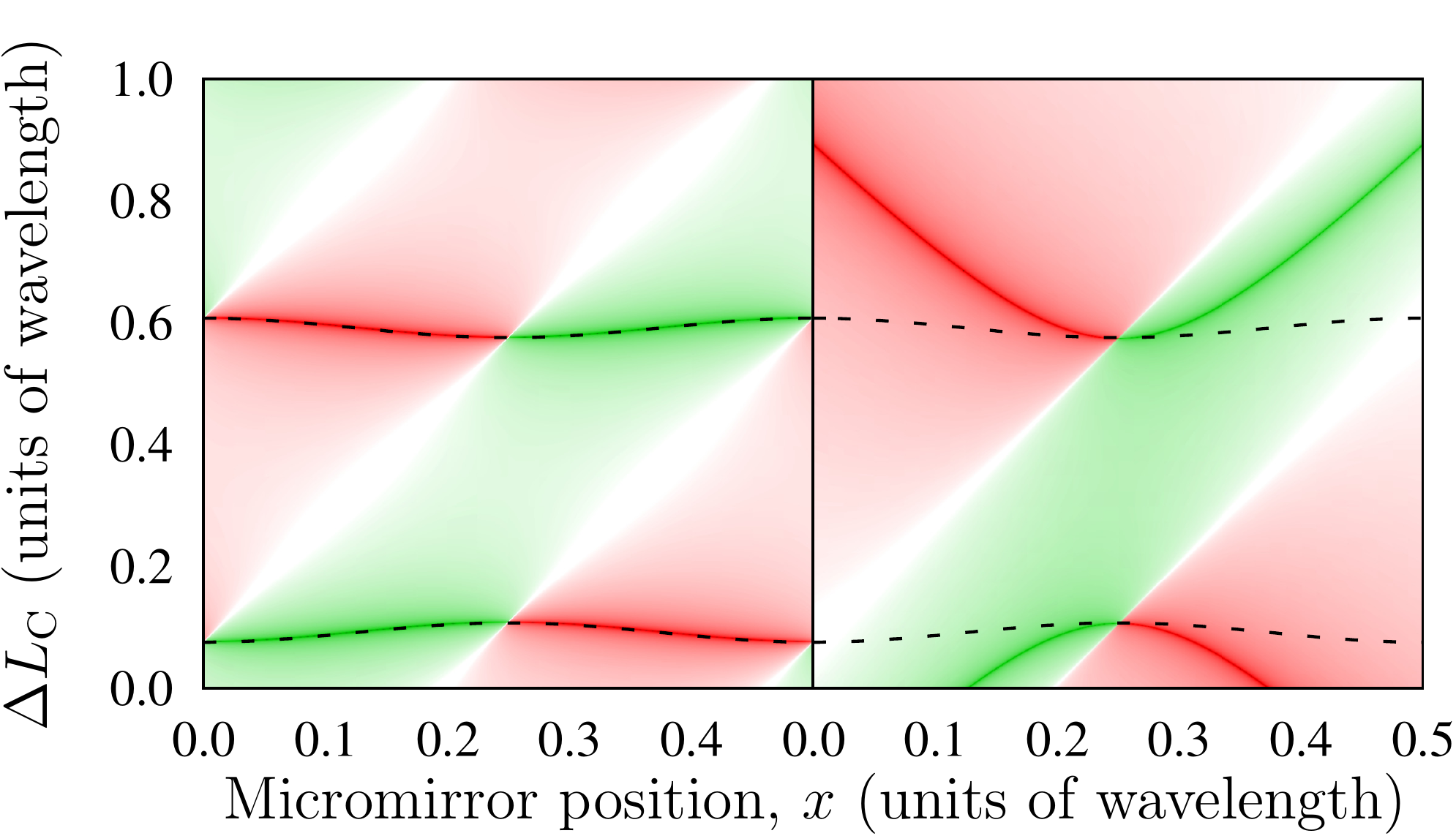}
 }
 \subfigure[\ $\zeta=-0.500$]{
  \includegraphics[angle=-90,width=0.4\textwidth]{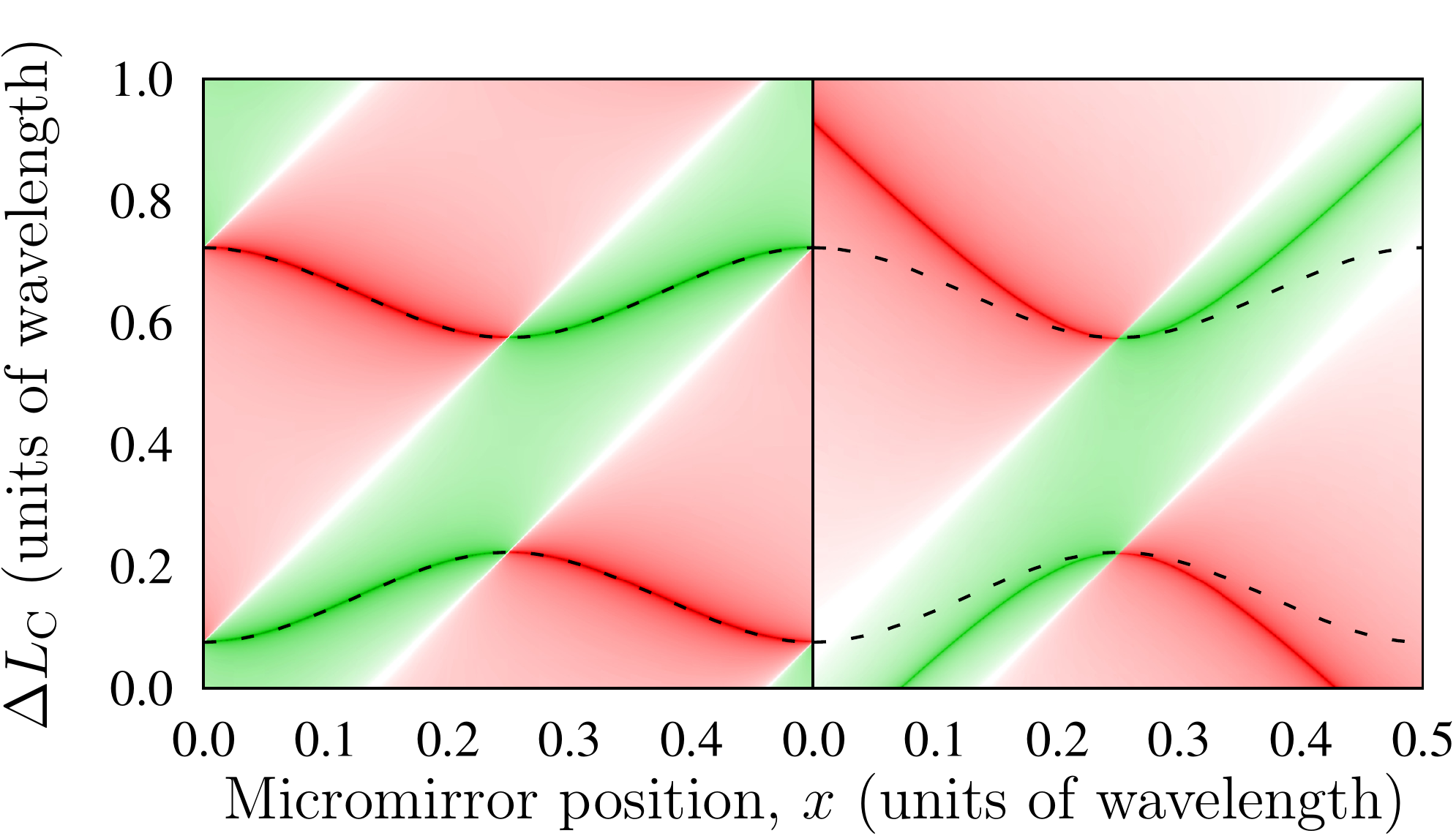}
 }\\
 \subfigure[\ $\zeta=-1.000$]{
  \includegraphics[angle=-90,width=0.4\textwidth]{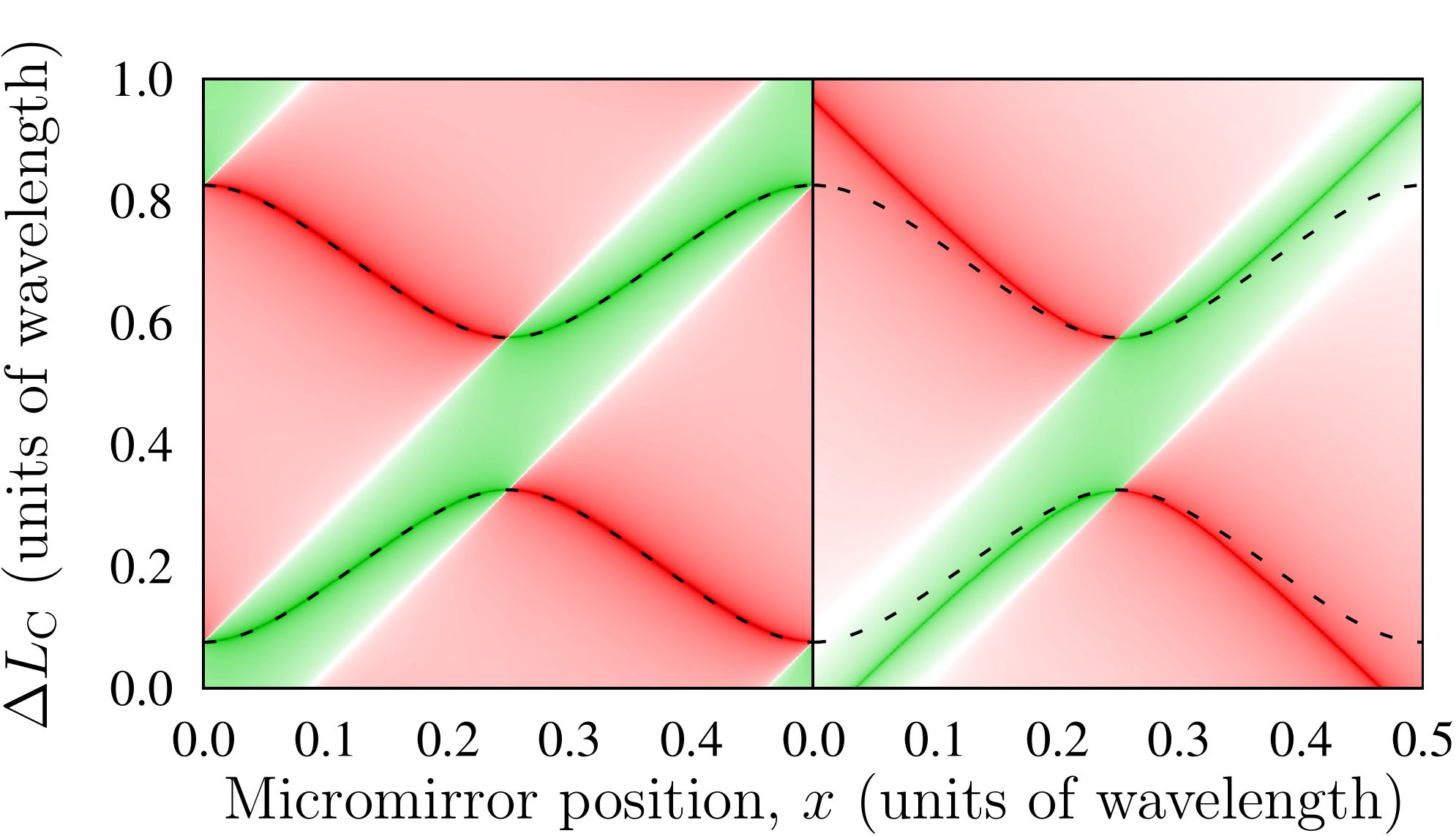}
 }
 \subfigure[\ $\zeta=-2.000$]{
  \includegraphics[angle=-90,width=0.4\textwidth]{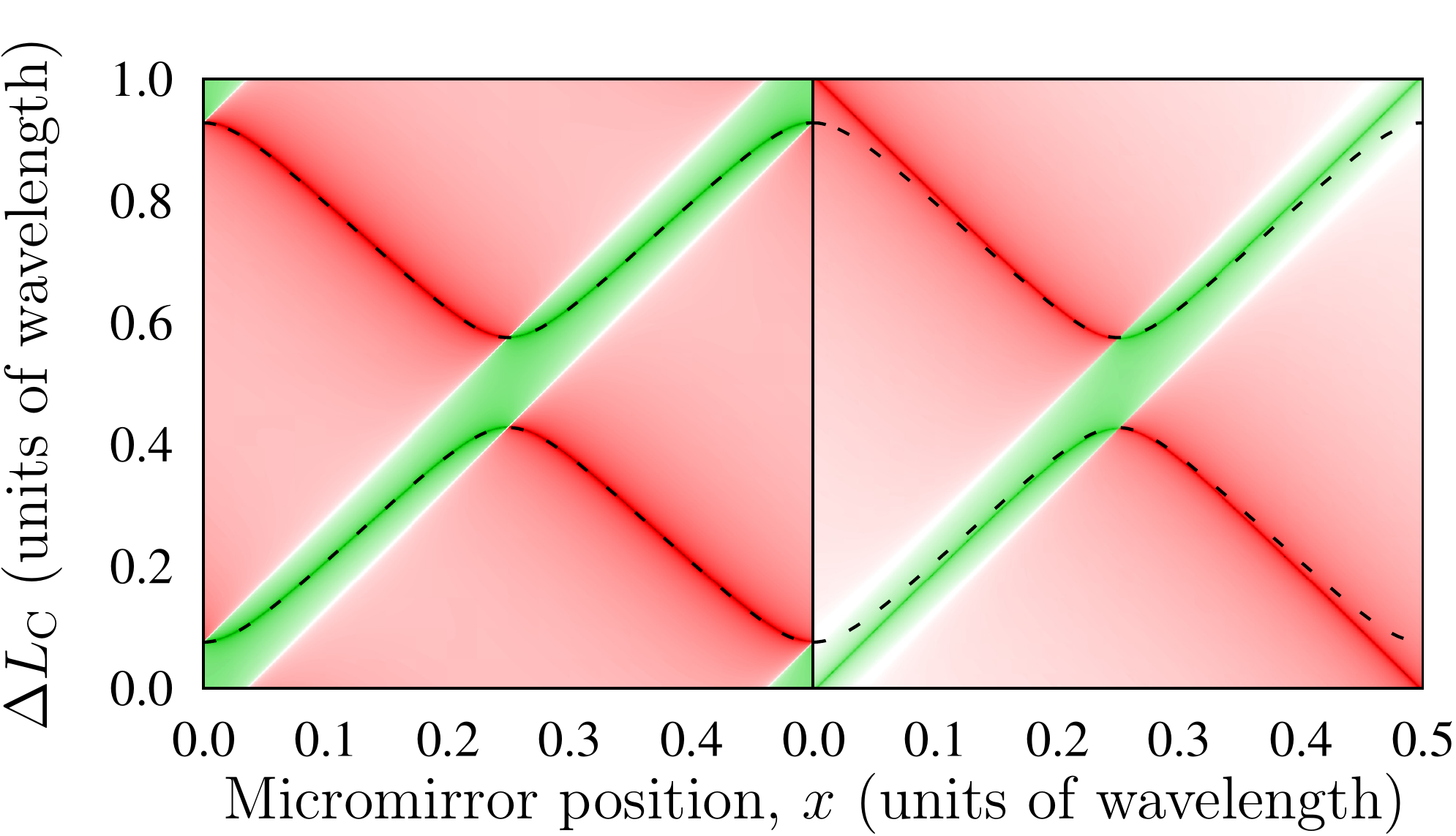}
 }\\
 \subfigure[\ $\zeta=-5.000$]{
  \includegraphics[angle=-90,width=0.4\textwidth]{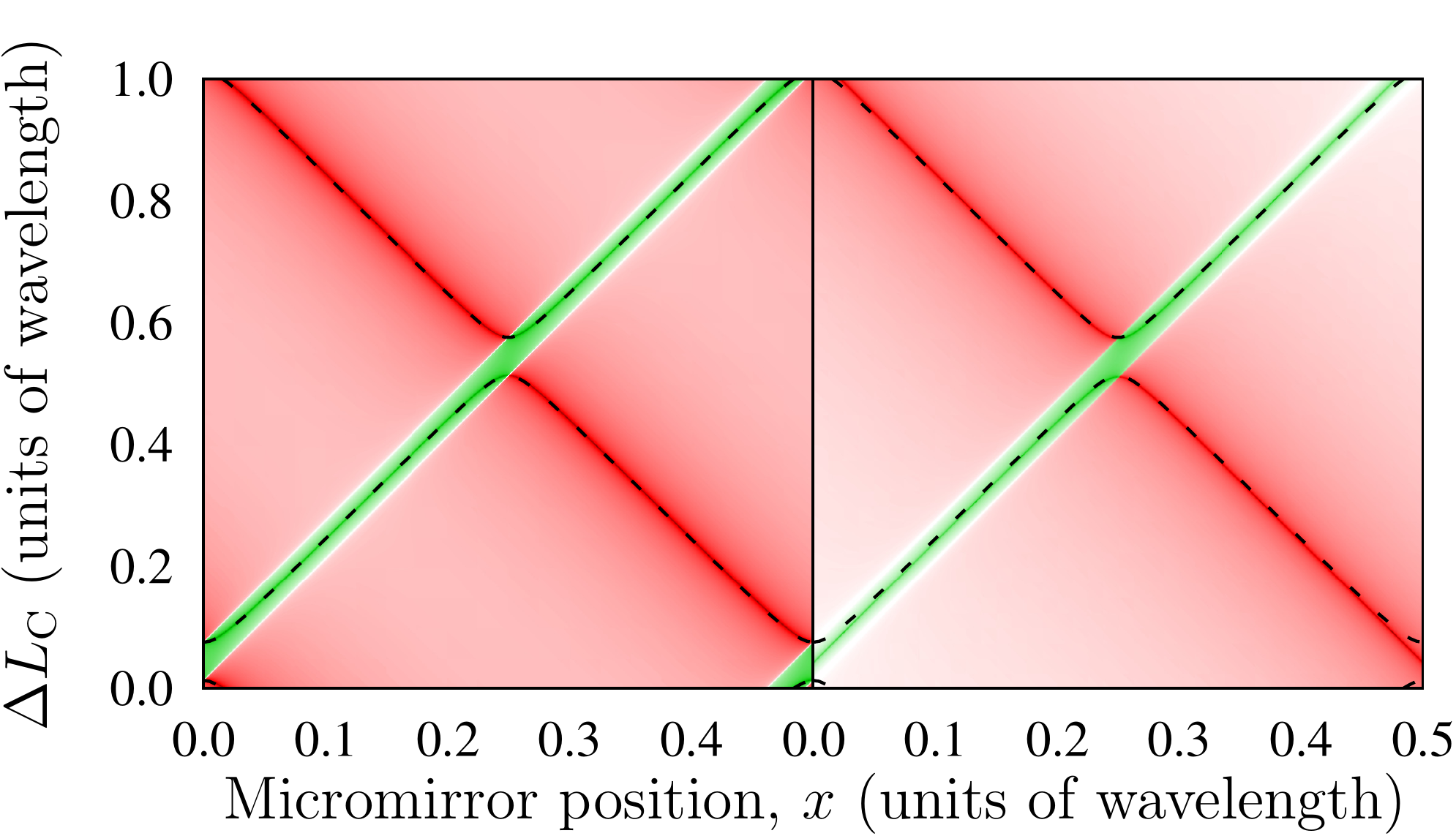}
 }
 \subfigure[\ $\zeta=-10.000$]{
  \includegraphics[angle=-90,width=0.4\textwidth]{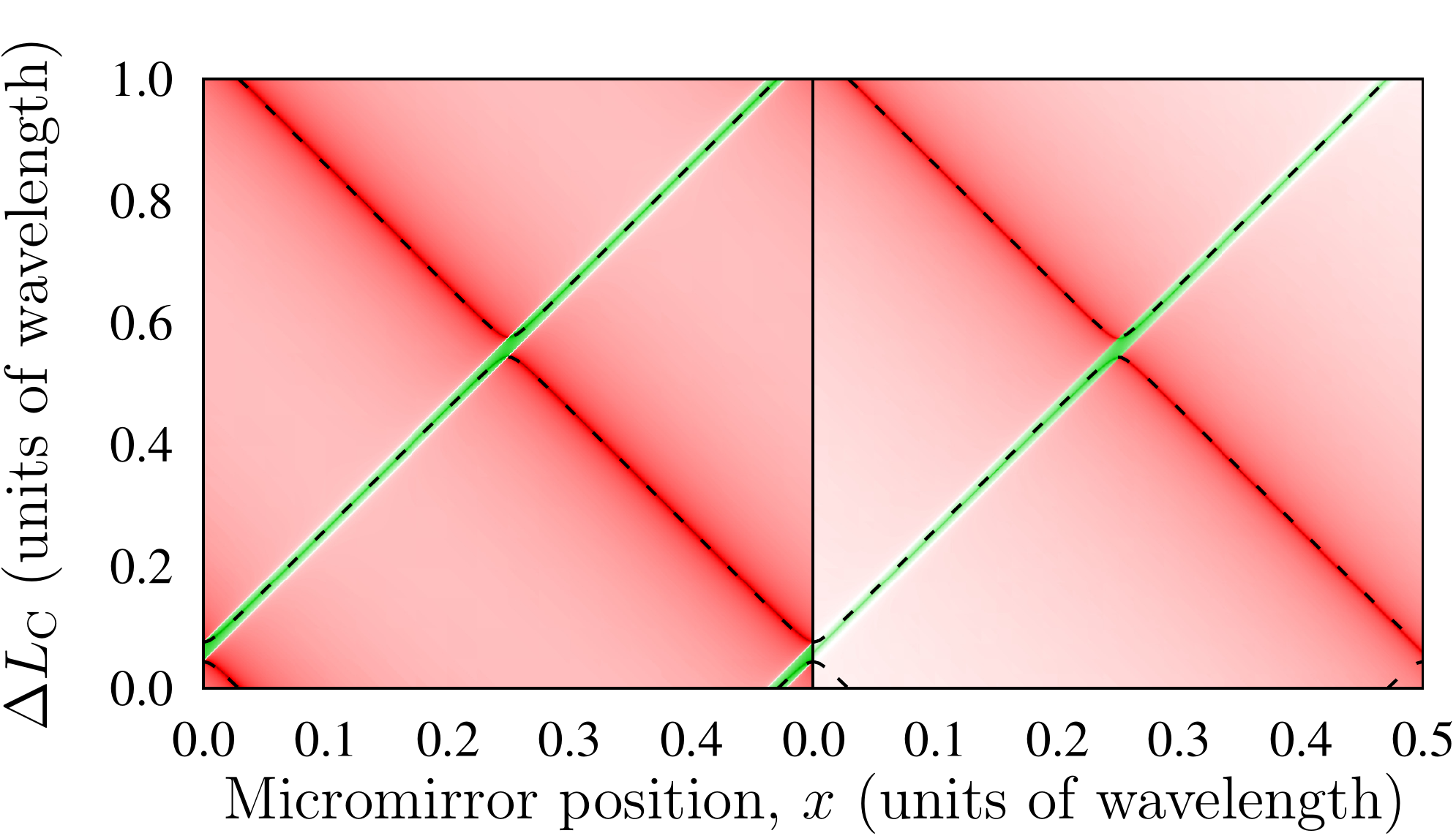}
 }\\
 \subfigure{
  \includegraphics[angle=-90,scale=0.3]{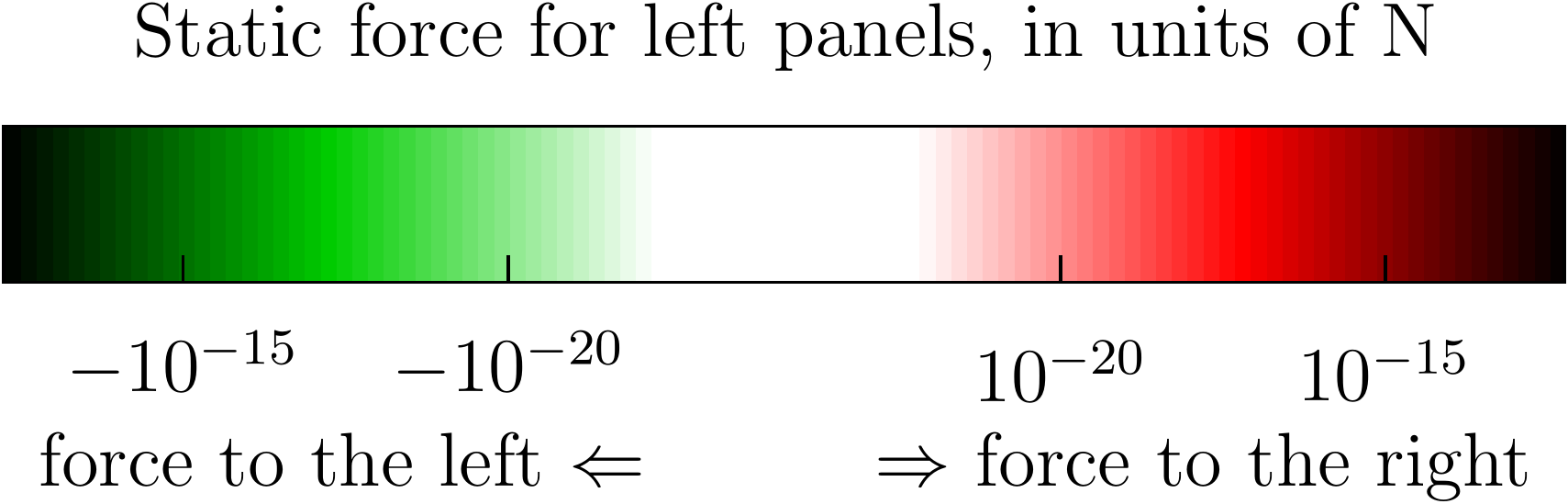}
 }
 \caption{Static force (\ie, the force acting on the mirror when $v=0$) computed from the scattering model presented here (left panels) and a model based on a modal decomposition~\cite{Jayich2008} (right panels), showing only one pair of modes. Red and green regions represent forces pointing in opposite directions, as indicated on the colourbar. We note qualitative agreement between the two models for $x\approx 0.25\lambda$ and for $\Delta L_\mathrm{c}$ close to the resonances, especially for large $\abs{\zeta}$. The discrepancies between the two sets of data, that are more pronounced for small polarisability, have significant consequences for any theory based on a coupled-cavity modal decomposition model. The black dashed lines [\eref{eq:CavityResonances}] trace the cavity resonances in the scattering model. The absolute values on the colourbar relate to the left panels.}
 \label{fig:SF}
\end{figure}

In parametrising our interaction in terms of a frequency shift $\Delta\omega$ we are effectively mapping the model originating from the TMM into a single-optical-mode model. It is important to note that this mode spans the entire cavity regardless of the nature of $\zeta$; what depends on $\zeta$ is the spatial profile of the mode. In the limit $\zeta\to0$, the field intensity is distributed uniformly throughout the cavity, whereas for large $\abs{\zeta}$, it is concentrated on one side of the membrane. These two situations are, as we have already discussed, handled differently in the CQED model, the former in terms of a single optical mode, and the latter in terms of two coupled optical modes. To highlight the failure of the coupled-optical-mode model as $\abs{\zeta}$ decreases, we show in \fref{fig:SF} the static force acting on the scatterer (\ie, the force when $v=0$) as predicted by the two models. For the coupled-mode model, we use the predictions of Ref.~\cite{Jayich2008}, which hold for $\abs{r}\to1$, and deliberately misapply them to cases where $\abs{r}\ll1$. From this model, given an input power $P_\mathrm{in}$, a tunnelling frequency $g=c\abs{t}/L_\mathrm{c}$, and a detuning $\Delta$ from resonance at $x=0$, one obtains
\begin{equation}
\force_0=-\frac{2\omega^\prime\kappa_\mathrm{c}}{k_0c}\frac{\kappa_\mathrm{c}^2+\bigl(\Delta+\omega^\prime x\bigr)^2-g^2}{(2\kappa_\mathrm{c}\Delta)^2+\bigl(\kappa_\mathrm{c}^2+\omega^{\prime\,2}x^2+g^2-\Delta^2\bigr)^2}P_\mathrm{in}\,,
\end{equation}
with $\omega^\prime=-2k_0c/L_\mathrm{c}$. For large $\abs{\zeta}$, the two descriptions are essentially identical; indeed, it is easy to understand that the description of two coupled cavities is a good one when the reflectivity of the central mirror approaches or exceeds $90$\%. For reflectivities of the order of $50$\% ($\abs{\zeta}\sim1$), however, the coupled-cavity description does not work well and one must switch to a scattering model to describe the situation accurately. For smaller $\abs{\zeta}$ still, as we have already mentioned, the predictions of the scattering model again agree with a CQED model of a scatterer (\eg, an atom) coupled to an unperturbed cavity field.

\section{Conclusion}
We have developed a generically-applicable theory to describe the motion of scatterers in electromagnetic fields. By applying this theory to the specific case of a scatterer in a cavity, we have shown how the scattering description can be used to bridge the gap between atom-CQED models, which rely on the atom interacting with one single mode that spans the entire cavity, and membrane-CQED models, where the membrane splits the cavity field into two coupled modes. It is in the region of current experimental interest, with membrane reflectivities of the order of $50$\%, that the discrepancy between the two descriptions starts emerging and where the usual ``$\abs{r}\to1$'' limit of membrane-CQED cannot be taken.

\section*{Acknowledgements}
This work was supported the Royal Commission for the Exhibition of 1851, the NSF (NF68736) and NORT (ERC\_HU\_09 OPTOMECH) of Hungary, and the UK EPSRC grants\linebreak EP/E058949/1 and EP/E039839/1.

\section*{References}
\bibliographystyle{iopart-num}
\providecommand{\newblock}{}

\end{document}